\documentstyle[epsfig,indentfirst,bm,12pt]{article}
\begin{document}
\title{\Large{\bf{The influence of quark energy loss on  extracting  nuclear sea quark distribution
from nuclear Drell-Yan experimental data }} }
\author{Duan Chun-Gui$^{1,2,4}$ \footnote{\tt{ E-mail:duancg$@$mail.hebtu.edu.cn}},
 Liu  Na$^{1,3}$}

\date{}

\maketitle \noindent {\small 1.Department of Physics, Hebei Normal
               University, Shijiazhuang 050016,China}\\
{\small 2.Hebei Advanced Thin Films Laboratory, shijiazhuang 050016,China}\\
{\small 3.College of Mathematics and Physics, Shijiazhuang University of Economics, Shijiazhuang 050031,China}\\
{\small 4.CCAST(World Laboratory), P.O.Box8730,Beijing 100080,China}

\baselineskip 9mm \vskip 0.5cm
\begin{abstract}

By means of two typical kinds of quark energy loss parametrization
and the nuclear parton distributions determined only with
lepton-nuclear deep inelastic scattering experimental data,  a
leading order analysis are performed on the proton-induced Drell-Yan
differential cross section ratios of tungsten versus deuterium as a
function of the quark momentum fraction in the beam proton and
target nuclei. It is found that the theoretical results with quark
energy loss are in good agreement with the experimental data.  The
quark energy loss effect produce approximately $3\%$ to $11\%$
suppression on the Drell-Yan differential cross section ratios
$R_{W/D}$ in the range $0.05\leq x_2\leq0.3$.  The application of
nuclear Drell-Yan data with heavy targets  is remarkably subject to
difficulty in the constraints of the nuclear sea-quark distribution.



\end{abstract}

\vskip 0.5cm

\hspace{0.5cm} After the discovery of the EMC effect$^{[1]}$, it is
found that the parton distributions are modified in the nuclear
environment. The origin of the nuclear effects is still under debate
in theory, and it is considered that different mechanisms are
responsible for the effects in the different regions of parton
momentum fraction. Up to now, almost all of the data on nuclear
dependence is from charged lepton deep inelastic scattering
experiments, which are sensitive only to the charge-weighted sum of
all quark and antiquark distributions. From the charged lepton deep
inelastic scattering off nuclei, the nuclear valence quark
distributions are relatively well determined in the medium and large
Bjorken $x$ regions. However, the charged lepton deep inelastic
scattering experiment would not be sensitive to the nuclear sea
quark distributions. In order to pin down the nuclear antiquark
distributions, it is desirable that the nuclear Drell-Yan reaction
is an ideal complementary tool.

\hspace{0.5cm}The production of large-mass lepton pairs from
hadron-hadron inelastic collisions was  first studied by Drell and
Yan, which is so-called Drell-Yan process$^{[2]}$. According to the
parton model, the process is induced by the annihilation of a
quark-antiquark pair into a virtual photon which subsequently decays
into a lepton pair. The nuclear Drell-Yan process of proton-nucleus
collisions therefore is closely related to the  quark distribution
functions in nuclei. It is further natural to expect that the
nuclear Drell-Yan reaction is a complementary tool to probe the
structure of hadron and nuclei.  The nuclear Drell-Yan experimental
data can be used to research the anti-quark distributions in hadron
and nuclei. However, it is indicated that the projectile rarely
retains a major fraction of its momentum in traversing the nucleus
in high energy inelastic proton-nucleus scattering. The quark and
gluon in the induced proton can loss a finite fraction of its energy
 due to the  multiple collisions in the nuclear target. In this respect, the
initial-state interactions should be important in nuclear Drell-Yan
process since the dimuon in the final state does not interact
strongly with the partons in the nuclei. The quark energy loss
effect in  nuclear Drell-Yan process is another nuclear effect apart
from the nuclear effects on the parton distribution as in deep
inelastic scattering.

\hspace{0.5cm} It is well known that the precise nuclear parton
distributions are very important to explain high energy nuclear
scattering phenomena. As for the strong necessities, the global
analysis of nuclear parton distribution functions have been proposed
in the recent years. For example, the global analysis of nuclear
parton distribution functions are given by EKRS(Eskola, Kolhinen,
Ruuskanen and Salgado)$^{[3]}$, HKN(Hirai, Kumano and Nagai)$^{[4]}$
and HKM (Hirai, Kumano and Miyama)$^{[5]}$. It is noticeable that
HKN employed Fermilab E772$^{[6]}$ and E866 $^{[7]}$ nuclear
Drell-Yan reaction data, EKRS included E772 experimental data,  and
HKM proposed the nuclear parton distributions which were determined
by means of the existing experimental data on nuclear structure
functions without including the proton-nucleus Drell-Yan process.

\hspace{0.5cm}Because of the quark energy loss effect in nuclear
Drell-Yan reaction, there is the kind of nuclear effect on the
differential cross section ratios as a function of the quark
momentum fraction in the beam proton and target nuclei. In previous
works$^{[8-11]}$, the Fermilab E866 $^{[7]}$ nuclear Drell-Yan
reaction data were used to investigate quark energy loss effect.
But, the impact of quark energy loss can be canceled partly out in
the proton-induced Drell-Yan cross section ratios for Fe to Be and W
to Be. In this paper, by combining two typical kinds of quark energy
loss parametrization with the nuclear parton distributions
determined only with lepton-nuclear deep inelastic scattering
experimental data, a leading order analysis are performed on the
differential cross section ratios of tungsten versus deuterium as a
function of the quark momentum fraction in the beam proton and
target nuclei for the proton-induced Drell-Yan process. The
influence of quark energy loss is elucidated on  extracting nuclear
sea-quark distribution.

\hspace{0.5cm}In the Drell-Yan process$^{[2]}$, the leading-order
contribution is quark-antiquark annihilation into a lepton pair. The
annihilation cross section can be obtained from the
$q\bar{q}\rightarrow\l^{+}l^{-}$ cross section, which is
\begin{equation}
  \sigma [q\bar{q}\rightarrow\l^{+}l^{-}]=\frac{4\pi\alpha_{em}^2}{9M^2}e^2_f,
\end{equation}
where $\alpha_{em}$ is the fine-structure constant, $e_f$ is the
charge of the quark and  $M$ is the invariant mass of the lepton
pair. The nuclear Drell-Yan differential cross section can be
written as
\begin{equation}
 \frac{d^2\sigma}{dx_1dx_2}=\frac{4\pi\alpha_{em}^2}{9sx_1x_2}
 \sum_{f}e^2_f[q^p_f(x_1,Q^2)\bar{q}^A_f(x_2,Q^2)
 +\bar{q}^p_f(x_1,Q^2)q^A_f(x_2,Q^2)],
\end{equation}
where$\sqrt{s}$ is the center of mass energy of the hadronic
collision,  $x_1$(resp.$x_2$)is the momentum fraction carried by the
projectile (resp.target) parton, the sum is carried out over the
quark flavor $f$, and $q^{p(A)}_{f}(x,Q^2)$ and ${\bar
q}^{p(A)}_{f}(x,Q^2)$ are respectively the quark and anti-quark
distributions in the proton (nucleon in the nucleus A).

\hspace{0.5cm}In order to take into account of the energy loss of
the fast quarks moving through the cold nuclei, we will introduce
two typical kinds of quark energy loss expressions, which are based
on the theoretical studies from Brodsky and Hoyer$^{[12]}$, and
Baier et.al.$^{[13]}$.  One can be rewritten as
\begin{equation}
\Delta x_1= {\alpha}\frac{<L>_A}{E_p},
\end{equation}
where $\alpha$ indicates the  incident quark energy loss per unit
length in nuclear matter, $<L>_A$ is the average path length of the
incident quark in the nucleus A, $E_p$ is the energy of the incident
proton. The average path length is employed by the conventional
value, $<L>_A=3/4(1.2A^{1/3)}$fm. In addition to the linear quark
energy loss rate, another is rewritten as
\begin{equation}
\Delta x_1= {\beta}\frac{<L>^2_A}{E_p}.
\end{equation}
Obviously, the partonic energy loss is quadratic with the path
length. With considering the  quark energy loss in nuclei, the
incident quark momentum fraction can be shifted from
$x'_1=x_1+\Delta x_1$ to $x_1$ at the point of fusion. Thus, the
nuclear Drell-Yan differential cross section can be expressed as
\begin{equation}
 \frac{d^2\sigma}{dx_1dx_2}=\frac{4\pi\alpha_{em}^2}{9sx_1x_2}
 \sum_{f}e^2_f[q^p_f(x'_1,Q^2)\bar{q}^A_f(x_2,Q^2)
 +\bar{q}^p_f(x'_1,Q^2)q^A_f(x_2,Q^2)].
\end{equation}
After the integral of the differential cross section above, the
nuclear Drell-Yan production cross section is given by
\begin{equation}
 \frac{d\sigma}{dx_{1(2)}}=\int dx_{2(1)}\frac{d^2\sigma}{dx_1dx_2}.
\end{equation}

\hspace{0.5cm}The Fermilab Experiment772(E772) $^{[6]}$ reported the
ratios of proton-induced nuclear  Drell-Yan differential cross
section for tungsten to deuterium target,
\begin{equation}
R_{W/D}(x_{1(2)})=\frac{d\sigma^{p-W}}{dx_ {1(2)}} /{\frac
{d\sigma^{p-D}}{dx_{1(2)}}}.
\end{equation}
Muon pairs were recorded in the range $4GeV\leq M \leq 9GeV$ and $M
\geq 11GeV$. The covered kinematical range was $0.1<x_2<0.3$. By
combining the quark energy loss effect with HKM $^{[5]}$ cubic type
of nuclear parton distributions determined only from lepton-nuclear
deep inelastic scattering experimental data, a global $\chi^2$
analysis is given to the experimental data on the ratios of nuclear
Drell-Yan differential cross section  $R_{W/D}(x_{1(2)})$.

\hspace{0.5cm}As  for the rations $R_{W/D}(x_{1})$,  the obtained
$\chi^2$ per degrees of freedom is $\chi^2/d.o.f.=2.14$ for the 31
data points without energy loss effect. It is shown that the
calculated results without energy loss effect are in disagreement
with the experimental data. With the fast quark energy loss, the
$\chi^2$ per degrees of freedom is  $\chi^2/d.o.f.= 0.84$ for the
linear quark energy loss formula with  $\alpha=1.28$, and the
$\chi^2$ per degrees of freedom is given by $\chi^2/d.o.f.= 0.84$
for the quadratic quark energy loss expression with $\beta=0.2$.  It
is found that the results given by the linear quark energy loss  are
nearly the same as these from the quadratic quark energy loss. The
nuclear Drell-Yan differential cross section ratios  are shown in
Fig.1 as functions of $x_1$ for various interval of $M$,
respectively. The solid curves are the ratios with only the nuclear
effects on the parton distribution. The dotted curves correspond to
the results by combining an linear quark energy loss with nuclear
effects on structure function. It is apparent that the results with
energy loss effect are in good agreement with the experimental data.

\hspace{0.5cm} For the nuclear Drell-Yan differential cross section
rations $R_{W/D}(x_{2})$, the $\chi^2$ per degrees of freedom is
given by $\chi^2/d.o.f.= 4.85$ for the 9 data points if the quark
energy loss is not put in. There  appears to be a significant
disagreement between the theoretical results without energy loss
effect and the experimental data. With considering the fast quark
energy loss, the $\chi^2$ per degrees of freedom is $\chi^2/d.o.f.=
1.31$ for the linear quark energy loss formula with $\alpha=0.94$,
and the $\chi^2$ per degrees of freedom is  $\chi^2/d.o.f.= 1.32$
for the quadratic quark energy loss expression with $\beta=0.15$. It
is shown that the results given by the linear quark energy loss  are
nearly identical with these from the quadratic quark energy loss. In
Fig.2, the solid curve is the ratios with only the nuclear effects
on the parton distributions, the dotted curve corresponds to the
results with adding linear energy loss. Compared with the
experimental data, our results with energy loss effect significantly
agree with the experimental data. Meanwhile, It is noticeable that
the values of the parameter $\alpha$ ( or $\beta$) are different by
means of the global $\chi^2$ analysis to the ratios $R_{W/D}(x_{2})$
and $R_{W/D}(x_{1})$. The results may be originated from the
experimental precision. If the experimental data are sufficiently
precise, the values of the parameter $\alpha$ (or $\beta$) in the
quark energy loss expression should be the same for fitting the
ratios $R_{W/D}(x_{2})$ and $R_{W/D}(x_{1})$ from the nuclear
Drell-Yan experiment.

\hspace{0.5cm} In order to clarify the influence of quark energy
loss on the Drell-Yan differential cross section ratios as a
function of the quark momentum fraction in target nuclei, the ratios
on $R_{W/D}(x_{2})$ without quark energy loss to those with linear
quark energy loss  are calculated and tabulated in Table 1.  The
similar results can be obtained for the quadratic quark energy loss.
It is found that the quark energy loss effect make  apparent
influence on the Drell-Yan differential cross section ratios
$R_{W/D}(x_{2})$.  The impact of quark energy loss on
$R_{W/D}(x_{2})$ become larger with the increase of momentum
fraction of the target parton.  The quark energy loss effect produce
approximately $3\%$ to $11\%$ suppression on the Drell-Yan
differential cross section ratios $R_{W/D}(x_{2})$ in the ranges
$0.05\leq x_2\leq0.3$.  Therefore, the application of nuclear
Drell-Yan data with heavy targets is remarkably subject to
difficulty in the constraints of the nuclear antiquark distribution.

\tabcolsep0.5cm
\begin{table}
\caption{The ratios of $R_{W/D}(x_{2})$ without quark energy loss to
those with linear quark energy loss.}
\begin{center}
\begin{tabular}{cccccccc}\hline
                   &{$x_2$}   & 0.05 & 0.10 & 0.15 & 0.20 & 0.25 & 0.30\\\hline
                   &the ratios of $R_{W/D}$  & 1.03 & 1.04 &1.05  & 1.07  & 1.09 & 1.11\\\hline
\end{tabular}
\end{center}
\end{table}

\hspace{0.5cm}In summary, the quark energy loss effect in  nuclear
Drell-Yan process is another nuclear effect apart from the nuclear
effects on the parton distribution as in deep inelastic scattering.
From the global $\chi^2$ analysis on the ratios of nuclear Drell-Yan
differential cross section $R_{W/D}(x_{1(2)})$, it is found that the
theoretical results with quark energy loss are in good agreement
with the experimental data. The quark energy loss effect has
significant impact on the Drell-Yan differential cross section ratio
as a function of the quark momentum fraction in the beam proton and
target nuclei. It is indicated that the influence of quark energy
loss on the differential cross section ratios become greater with
the increase of momentum fraction of the target parton. The
application of nuclear Drell-Yan data with heavy targets  is
remarkably subject to difficulty in the constraint of the nuclear
antiquark distribution. Therefore, we desire to operate precise
measurements of the experimental study from the relatively low
energy nuclear Drell-Yan process at J-PARC$^{[14]}$ and Fermilab
E906$^{[15]}$. These new experimental data on nuclear Drell-Yan
reaction can shed light on the energy loss of fast quark propagating
in a cold nuclei, and the impact of quark energy loss on
determination nuclear sea-quark distribution .

{\bf Acknowledgement:} This work is partially Supported  by National
Natural Science Foundation of China (10575028), Natural Science
Foundation of Hebei Province (A2008000137).

\newpage

\begin{figure}
\centering
\includegraphics[width=1.0\textwidth]{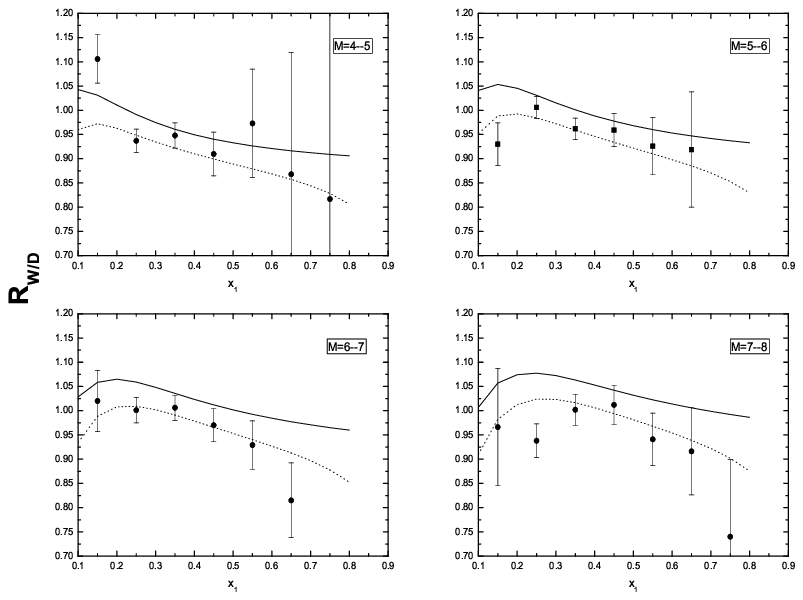}
\caption{ The nuclear Drell-Yan cross section ratios $R_{W/D}(x_1)$
on for various intervals M. Solid curves correspond to nuclear
effects on structure function. Dotted curves  show the combination
of linear quark energy loss effect with HKM cubic type of nuclear
parton distributions. The experimental data are taken from
E772$^{[6]}$.}
\end{figure}

\begin{figure}
\centering
\includegraphics[width=1.0\textwidth]{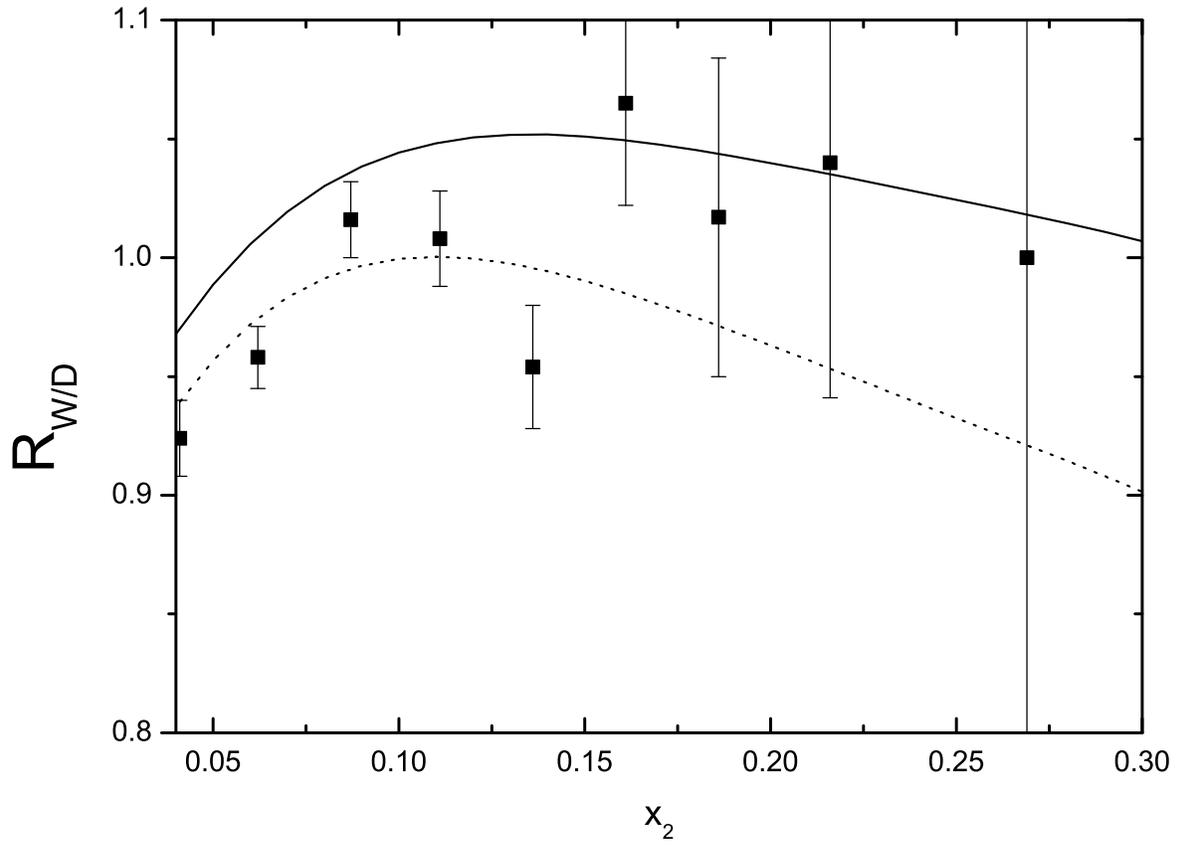}
\caption{The nuclear Drell-Yan cross section ratios $R_{W/D}(x_2)$
of tungsten versus deuterium. The comments are the same as Fig.1.}
\end{figure}

\end{document}